\documentclass[cits]{PoS}

\title{{\it Swift} monitoring of Supergiant Fast X--ray Transients: the 
out-of-outburst behaviour and the flares from IGR~J17544--2916 and 
XTE~J1739--302}

\ShortTitle{{\it Swift} monitoring of SFXTs}

\author{ \speaker{L.\ Sidoli},$^a$ P.\ Romano,$^b$  G.\ Cusumano,$^b$ 
V.\ Mangano,$^b$ S.\ Vercellone,$^a$ A.\ Paizis,$^a$ A.\ Pellizzoni,$^c$ 
J.A.\ Kennea,$^d$ D.N.\ Burrows,$^d$
H.A.\ Krimm,$^{efg}$  N.\ Gehrels,$^g$ 
C.\ Guidorzi,$^h$  P.A.\ Evans$^i$  \\
\llap{$^a$}INAF, Istituto di Astrofisica Spaziale e Fisica Cosmica, \\
         Via E.\ Bassini 15,   I-20133 Milano,  Italy\\
\llap{$^b$}INAF, Istituto di Astrofisica Spaziale e Fisica Cosmica, \\
         Via U.\ La Malfa 153, I-90146 Palermo, Italy\\
\llap{$^c$}INAF, Osservatorio Astronomico di Cagliari, \\
         Localit\`a Poggio dei Pini, strada 54, I-09012 Capoterra, Italy \\ 
\llap{$^d$}Department of Astronomy and Astrophysics, Pennsylvania State  University, \\
         University Park, PA 16802, USA\\
\llap{$^e$}CRESST/Goddard Space Flight Center, Greenbelt, MD, USA\\
\llap{$^f$}Universities Space Research Association, Columbia, MD, USA\\
\llap{$^g$}NASA/Goddard Space Flight Center, Greenbelt, MD 20771, USA\\
\llap{$^h$}INAF, Osservatorio Astronomico di Brera, \\ 
           via E.\ Bianchi 46, I-23807 Merate, Italy\\
\llap{$^i$}Department of Physics \& Astronomy, University of Leicester, LE1 7RH, UK\\
E-mail: \email{sidoli@iasf-milano.inaf.it}
}
\abstract{Supergiant Fast X--ray Transients (SFXTs) are a sub-class of High Mass X--ray Binaries (HMXBs) 
associated with OB supergiant companions and displaying transient 
X--ray activity.
This behaviour is quite surprising since HMXBs 
hosting supergiants were known to be persistent sources, until the INTEGRAL discoveries 
obtained by means of the monitoring of the Galactic plane.
We have been performing a monitoring campaign with {\it Swift} of four SFXTs with the main aim of characterizing
both the long-term behaviour  of these transients and the properties during bright outbursts.
Here we discuss the properties of the X--ray emission observed outside the outbursts
as well as the flares observed from two SFXTs: IGR~J17544--2916 and XTE~J1739--302.
Contrarily to what previously thought, \ {\it Swift} allowed us to discover that 
SFXTs spend most of the time in accretion at a low level, even outside the bright outbursts,
with an accretion luminosity of 10$^{33}$--10$^{34}$~erg~s$^{-1}$, and that the quiescent level 
$\sim$10$^{32}$~erg~s$^{-1}$, is a much rarer state.
}

\FullConference{7th INTEGRAL Workshop\\
		 September 8-11 2008\\
		 Copenhagen, Denmark}

\begin{document}

\section{Supergiant Fast X--ray Transients before {\it Swift}}

The Galactic plane monitoring performed with the INTEGRAL satellite led to
the discovery of several new sources (Bird et al., 2007). 
Some of them displayed  
sporadic, recurrent, bright and short flares, with a typical duration of a few hours and reaching
a peak luminosity of 10$^{36}$--10$^{37}$~erg~s$^{-1}$
(Sguera et al, 2005, 2006; Negueruela et al. 2006).
Refining the INTEGRAL positions at arcsec level with 
X--ray follow-up observations, allowed the 
association with OB supergiant companions
(e.g. Halpern et al.\ 2004; Pellizza et al.\ 2006; Masetti et al.\ 2006;
Negueruela et al.\ 2006b; Nespoli et al.\ 2008).

Other important properties are the spectral similarity with 
accreting pulsars (hard power law spectra with a high energy cut-off around 15--30~keV) 
and the large dynamic range, from a peak luminosity
of 10$^{36}$--10$^{37}$~erg~s$^{-1}$, down to a quiescent emission of  10$^{32}$~erg~s$^{-1}$.
The two main characterizing properties (the transient X--ray emission and
the association with  supergiant companions) 
indicate that these transients form a new class of High Mass X--ray Binaries, 
later called
Supergiant Fast X--ray Transients (SFXTs; e.g. Negueruela et al. 2006).

The similarities of the SFXTs with the properties of accreting pulsars suggest 
that the majority of these transients are indeed HMXBs hosting a neutron star,
although only in three SFXTs X--ray pulsations have been discovered: 
IGR~J11215--5952 ($P_{\rm spin}$$\sim$186.8 \,s, Swank et al.\ 2007); 
AX~J1841.0--0536 ($P_{\rm spin}$$\sim$4.7\,s, Bamba et al.\ 2001)
and  IGR~J18483--0311 ($P_{\rm spin}$$\sim$21 \,s, Sguera et al.\ 2007).

The confirmed SFXTs are eight 
(IGR~J08408--4503, IGR~J11215--5952, IGR~J16479--4514, XTE~J1739--302, IGR~J17544--2619,
SAX~J1818.6--1703, AX~J1841.0-0536 and IGR~J18483--0311),
with $\sim$15 more candidates which
showed short transient flaring activity,
but with no confirmed association with an OB supergiant companion.

The main mechanisms proposed to explain the short and bright flaring activity
from SFXTs deal with the properties of the accretion from the 
supergiant wind (see Sidoli 2008 for a review), either 
related with the wind structure  (in't Zand 2005; Walter \& Zurita Heras, 2007;
Negueruela et al. 2008; Sidoli et al. 2007) or to gated mechanisms which allow accretion onto
the neutron star surface only when the centrifugal or the magnetic barriers are open, depending
on the values of the neutron star spin and surface magnetic field (e.g. Bozzo et al. 2008 and references therein).

The properties of the SFXTs outbursts, although sporadic and short, 
have been studied more in depth than the quiescent state.
The observations performed outside the bright outbursts have been indeed only a few and short (a few ks long), 
and caught these sources either in a low level flaring activity (IGR~J17544--2619, Gonzalez-Riestra et al. 2004) 
or in quiescence (with a very soft spectrum, likely thermal, with an 
X--ray luminosity of $\sim$10$^{32}$~erg~s$^{-1}$). 
Note that this latter quiescent state has been observed
{\em only} in a couple of SFXTs, IGR~J17544--2619 (in't Zand 2005) and IGR J08408--4503
(Leyder et al.\ 2007).

\begin{figure}[ht!]
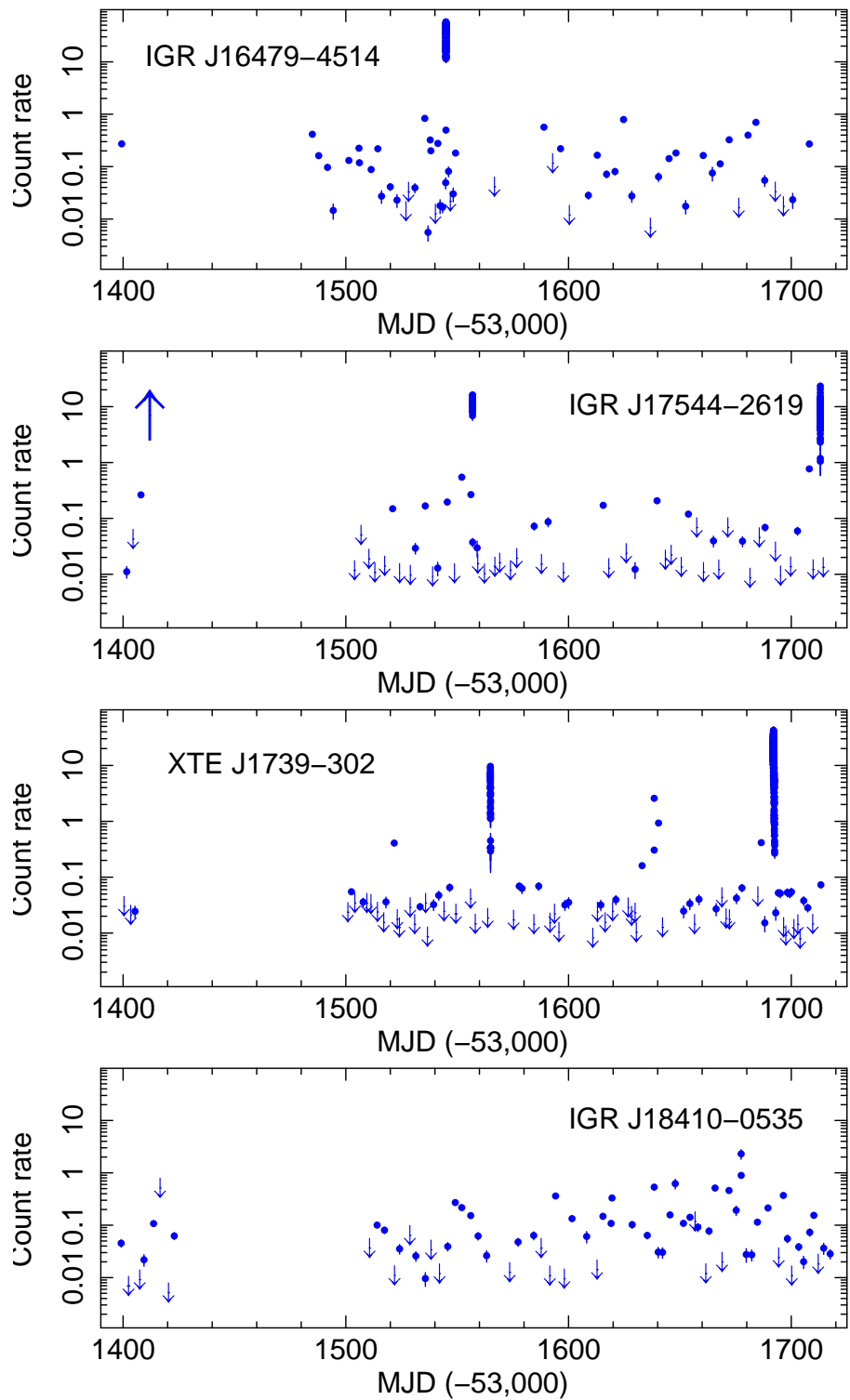

\begin{center}
\includegraphics*[angle=270,scale=0.5]{inte_igr16479_lc.ps}
\includegraphics*[angle=270,scale=0.5]{inte_igr17544_lc.ps}\\
\includegraphics*[angle=270,scale=0.5]{inte_xte1739_lc.ps}
\includegraphics*[angle=270,scale=0.5]{inte_igr18410_lc.ps}\\
\end{center}
\caption{\scriptsize Light curves of the 4 SFXTs monitored with \emph{Swift}/XRT (0.2--10 keV), 
from 2007 October to 2008 September 10. 
The upward pointing arrow in the IGR~J17544--2619 light curve 
marks an outburst which triggered the BAT Monitor 
on MJD 54412  (2007-11-08) but could not be observed with XRT because the source was 
Sun-constrained. The downward-pointing arrows are 3-$\sigma$ upper limits. 
The  gap in the observations between about December 2007 and January 2008
is because the sources were Sun-costrained.
}
\label{lsfig:4lc}
\end{figure}

\section{{\it Swift} monitoring of Supergiant Fast X--ray Transients}

Before the {\it Swift} campaign (which is still in progress since October 2008) 
no long-term monitoring of SFXTs have ever been performed to study 
the status where these transients spend most of their life. 
Nevertheless, it has been assumed by several authors, without observational evidence, that 
SFXTs spend most of the time in quiescence, when they are not in bright outburst.

The first observations with {\it Swift} of a member of this new class of sources 
have been performed during the 2007 February outburst of IGR~J11215--5952 (Romano et al. 2007).
This outburst could be completely monitored thanks to its predictability, because  IGR~J11215--5952 was
the first SFXT where periodically recurrent outbursts were discovered (Sidoli et al. 2006).
These observations are one of the most complete set of observations of a SFXT in outburst,
and clearly demonstrate, for the first time, that the short (a few hours long) 
flares observed with INTEGRAL (or RXTE, in a few  sources), are actually 
part of a much longer outburst event lasting a few days, 
implying that the accretion phase lasts longer than what was previously thought 
(Romano et al. 2007; Sidoli et al. 2007).

The success of this campaign led us to propose with {\it Swift} the first wide-band,  long-term 
and deep monitoring campaign of a sample of four SFXTs,
with the main aim of
(1)-studying the long-term properties of these transients, (2)-performing 
a truly simultaneous spectroscopy 
(0.3--150 keV) during outbursts, (3)-studying the outburst recurrence and their durations (see 
also Romano et al. 2008b, these proceedings).
The 4 targets are: XTE~J1739--302, IGR~J17544--2619, IGR~J16479--4514
and AX~J1841.0--0536/IGR~J18410--0535.
The {\it Swift} campaign consists of 2--3 observations/week/source (each observation lasts 1--2~ks; 
see  Romano et al. 2008b, these proceedings, for the campaign strategy).
Fig.~\ref{lsfig:4lc} shows the four  {\it Swift}/XRT light curves (0.2--10~keV) 
accumulated in the period October 2007--September 2008.

Here we report on the entire {\it Swift} monitoring campaign, updated to 2008 September 10.
In particular, we  focus on the out-of-outburst behaviour 
(Sidoli et al. 2008a, hereafter Paper~I) 
and on the bright flares observed 
from two SFXTs of the sample, XTE~J1739--302 and IGR~J17544--2619 
(Sidoli et al. 2008b, hereafter Paper~III; Sidoli et al. in preparation). 
Another outburst caught during this campaign
from IGR~J16479--4514 was published by Romano et al. (2008a, Paper~II).

Preliminary results from the last outbursts from XTE~J1739--302 (triggered on 2008 August 13, Romano et al. 2008c) 
and from IGR~J17544--2619 (triggered on 2008 September 4,  Romano et al. 2008d) 
are also discussed here for the first time. 
A complete analysis will be addressed in Sidoli et al. (in preparation).

\subsection{SFXTs: the long-term  X-ray emission outside the bright outbursts}

The SFXTs light curves of Fig.~\ref{lsfig:4lc} show a clear 
evidence for highly variable source fluxes even outside the bright outbursts (which were caught
in three of the four sources we are monitoring).
The light curve variability is on  timescales of days,
weeks and months, with a dynamic range (outside bright outbursts)
of more than one order of magnitude in all four SFXTs.
These sources spend most of the time in a 
frequent low-level flaring activity
with an average 2--10 keV luminosity of about 10$^{33}$--10$^{34}$~erg~s$^{-1}$ (see Paper~I).

The average spectra of this out-of-outburst emission are hard (although not as hard as during the
bright flares) and can be fitted with an absorbed power law
with a photon index in the range 1--2. The absorbing column density is typically higher than
the Galactic value, which can be derived from the optical extinction toward the optical counterparts.

The out-of-outburst emission in IGR~J16479--4514  and in AX~J1841.0--0536 appears to be modulated
with a periodicity in the range 22--25~days, although a full timing analysis
will be addressed at the end of the campaign.
The spectral properties together with the high dynamic range in the flux 
variability when the sources are {\em not} in outburst, demonstrate 
that SFXTs still accrete matter even outside their bright outbursts,
and that the quiescent state (characterized by a very soft spectrum and by a low level of emission
at about 10$^{32}$~erg~s$^{-1}$) is not the typical long-term state in SFXTs.

\subsection{SFXTs: bright flares from IGR~J17544--2619 and XTE~J1739--302}

Typically, the SFXTs long-term light curves show a number of bright outbursts, reaching peak luminosities 
of a few 10$^{36}$~erg~s$^{-1}$, assuming the  distances determined by Rahoui et al. (2008).
The only source which did not undergo  bright flares is AX~J1841.0--0536/IGR~J18410--0535,
which  showed a flux variability of more than two orders of magnitude.

During the {\it Swift} campaign, three and two outbursts were caught respectively from 
IGR~J17544--2619 (the first of them triggered BAT, but could not be observed with {\it Swift}/XRT 
because of Sun-constraints) and from XTE~J1739--302,  at the following
dates: on 2007 November 8, 2008 March 31 and 2008 September 4 from IGR~J17544--2619, and
on 2008 April 8 and 2008 August 13 from XTE~J1739--302.
Thus, bright flares in this two prototypical SFXTs occur on a timescale of $\sim$4--5 months
(the three outbursts from IGR~J17544--2619 were spaced by $\sim$144 and 157 days, respectively, while
the two outbursts from XTE~J1739--302 were spaced by 127~days).

The bright flare from IGR~J17544--2619 (on 2008 March 31; Paper~III) 
could be observed simultaneously with XRT (0.2--10~keV) and BAT (15--150~keV).
A fit with a power law with a high energy cut-off ($e^{(E_{\rm cut}-E)/E_{\rm fold}}$) 
resulted in the following parameters:
$N_{\rm H}$=(1.1$\pm{0.2}$)$\times 10^{22}$~cm$^{-2}$, $\Gamma$=0.75$\pm{0.11}$,    
cut-off energy $E_{\rm cut}$=18$\pm{2}$~keV
and e-folding energy $E_{\rm fold}$=4$\pm{2}$~keV, reaching a 
luminosity of  5$\times$10$^{37}$~erg~s$^{-1}$ (0.5--100~keV at 3.6~kpc).
Note that the out-of-outburst emission observed with XRT below 10 keV
is softer and more absorbed than the emission during this flare.

The other flare observed from IGR~J17544--2619 on 2008 September 4 was even brighter (Romano et al. 2008d), 
and was preceeded by intense activity for a few days  as observed with INTEGRAL 
during the Galactic bulge monitoring programme (Kuulkers et al. 2008; Romano et al. 2008d).
The XRT light curve exceeded 20~s$^{-1}$. 
This peak emission
could be fitted with an absorbed power law with a photon index of 1.3$\pm{0.2}$ and 
an absorbing column density of 1.8$^{+0.4}_{-0.3}$$\times10^{22}$~cm$^{-2}$.
The average flux in the 2--10~keV range was  8$\times$10$^{-10}$~erg~cm$^{-2}$~s$^{-1}$.
The fainter X--ray emission during the flare (2$\times$10$^{-10}$~erg~cm$^{-2}$~s$^{-1}$) 
displayed a similar 
absorbing column density of 1.4$^{+0.7}_{-0.5}$$\times10^{22}$~cm$^{-2}$ and a photon index 
$\Gamma$=0.8 $^{+0.4}_{-0.3}$.
A more detailed analysis of the properties of this outburst will be performed in a
forthcoming paper (Sidoli et al. in preparation).

The first outburst from XTE~J1739--302 was caught on 2008 April 8 (Paper~III) and was composed 
by two bright flares separated by about 6000~s.
The X--ray emission was significantly  more absorbed than in   IGR~J17544--2619:
the broad band (XRT+BAT) spectrum could be well described by an
absorbed  high energy cut-off 
power law with the following parameters: $N_{\rm H}$=1.3$\times$$10^{23}$~cm$^{-2}$, 
$\Gamma$=1.4$^{+0.5} _{-1.0}$,
cut-off energy $E_{\rm cut}$=6$ ^{+7} _{-6}$~keV
and e-folding energy $E_{\rm fold}$=16 $ ^{+12} _{-8}$~keV.
The derived X--ray luminosity is  3$\times$10$^{37}$~erg~s$^{-1}$ (0.5--100~keV).

\begin{figure}[ht!]
\begin{center}
\includegraphics*[angle=270,scale=0.45]{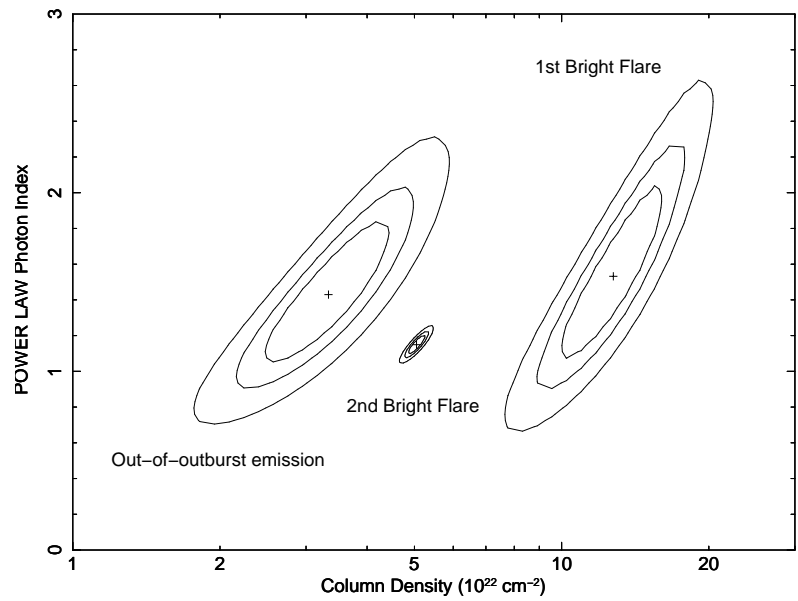}
\end{center}
\caption{\scriptsize Comparison of the spectral paramenters (absorbed single power law model)
derived for  XTE~J1739--302 during the two bright flares discussed here, 
and the total spectrum of the out-of-outburst emission  reported in Paper~I.
68\%, 90\% and 99\% confidence level contours are shown.
}
\label{lsfig:contxte}
\end{figure}

A new outburst was caught from XTE~J1739--302 on 2008 August 13 (Romano et al. 2008c).
A preliminary spectral analysis of the average broad band spectrum of this bright flare resulted
in the following parameters, adopting an absorbed power law with a high energy cut-off:
absorbing column density $N_{\rm H}$=(4.0$\pm{0.3}$)$\times$$10^{23}$~cm$^{-2}$, 
$\Gamma$=0.7$\pm{0.1}$,
$E_{\rm cut}$=4.6$\pm{0.3}$~keV
and $E_{\rm fold}$=9 $ ^{+2} _{-1}$~keV. 
The X--ray luminosities during the flare were 2$\times$10$^{36}$~erg~s$^{-1}$ (0.5--10~keV) and
5$\times$10$^{36}$~erg~s$^{-1}$ (0.5--100~keV). 
Fig.~\ref{lsfig:contxte} shows the comparison of the spectroscopy in the
soft energy range (XRT data) of the out-of-outburst emission with the results
from the two flares from XTE~J1739--302.
A time resolved spectral 
analysis during the flare will be reported in a forthcoming paper (Sidoli et al., in preparation).

\begin{figure}[ht!]
\begin{center}
\includegraphics*[angle=0,scale=0.7]{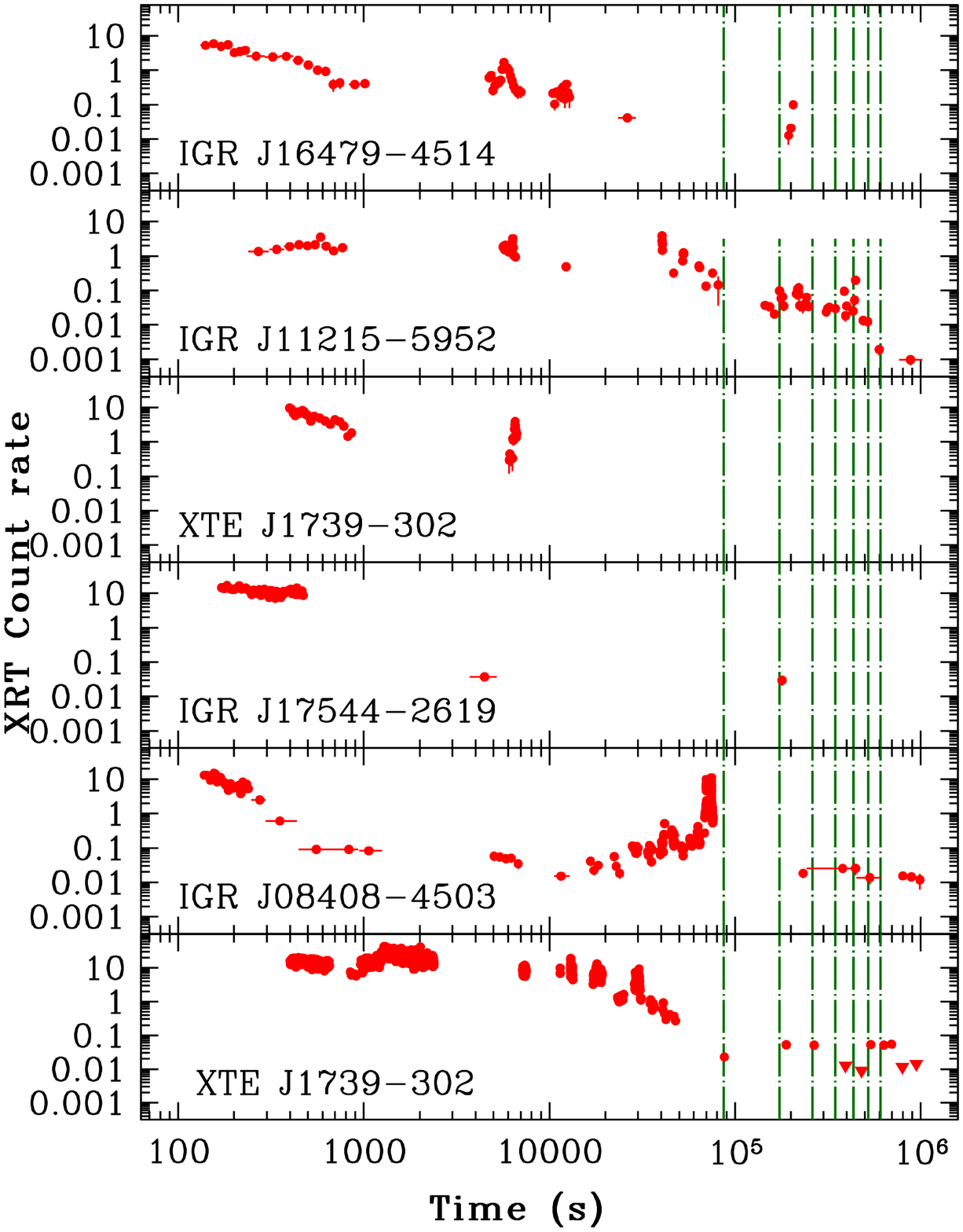}
\end{center}
\caption{\scriptsize Light curves of the outbursts of SFXTs followed by {\it Swift}/XRT
referred to their
respective triggers. We show the 2005 outburst of IGR~J16479$-$4514 (Paper~I), 
which is more complete than the one observed in 2008 (Paper~II).
The IGR~J11215$-$5952 light curve has an arbitrary start time, since
the source
did not trigger the BAT (the observations were obtained as a ToO; Romano et al. 2007).
The third and the last panels report the two flares from  XTE~J1739--302 observed
on 2008 April 8 and on 2008 August 13, respectively.
The forth panel shows the outburst from IGR~J17544--2916 occurred on 2008 March 31 (Paper~III).
The fifth panel reports on a multiple flaring activity reported from another SFXT,
not part of this campaign, IGR~J08408--4503, and occurred on 2008 July 5 (Romano et al., 2008e).
Note that where no data are plotted, no data were collected. Vertical dashed lines 
mark time intervals equal to 1 day.
}
\label{lsfig:duration}
\end{figure}

A comparison of the SFXTs light curves (the four SFXTs constantly monitored with {\it Swift},
together with other two sources, IGR~J11215--5952 and IGR~J08408--4503) 
during their outbursts  are reported in
Fig.~\ref{lsfig:duration}.
This plot clearly demonstrates that the outbursts from 
all these transients last much longer than simply a few hours as previously thought.
Fig.~\ref{lsfig:duration} shows about 8 days of monitoring for each target, and 
it is clear that the first SFXT, where a day-long outburst event has been observed
(IGR~J11215--5952, Romano et al. 2007), is not a peculiar case among SFXTs, but a similar behaviour
has been observed in the other SFXTs monitored by {\it Swift} during the last year 
(except AX~J1841.0--0536, where no outburst have yet been observed).

\section{Conclusions}

The results of the monitoring campaign we have been performing in the last year 
with {\it Swift} of a sample of 4 SFXTs can be summarized as follows:

\begin{itemize}

\item the long-term behaviour of the SFXTs outside their outbursts is a low-level accretion phase at a
luminosity of 10$^{33}$--10$^{34}$~erg~s$^{-1}$, with a dynamic range of 1 up to, sometimes, 2 orders of magnitude in flux;

\item the broad band X--ray emission during the bright flares can be described well with models 
commonly adopted for the emission from the accreting X--ray pulsars;

\item the SFXTs spectra during flares show high energy cut-offs 
compatible with a neutron star magnetic field of about 10$^{12}$~G, although no cyclotron lines have been detected yet;

\item the duration of the outbursts from different SFXTs observed with {\it Swift} are longer than a few hours.

\end{itemize}

\acknowledgments
We thank the {\it Swift} team duty scientists and science planners  P.J.\ Brown, M.\ Chester,
E.A.\ Hoversten, S.\ Hunsberger,  C.\ Pagani, J.\ Racusin, and M.C.\ Stroh
for their dedication and willingness to accomodate our sudden requests
in response to outbursts during this long monitoring effort.
We also thank the remainder of the {\it Swift} XRT and BAT teams,
J.A.\ Nousek and S.\ Barthelmy in particular, for their invaluable help and support with
the planning and execution of the observing strategy.
This work was supported in Italy by contracts ASI I/023/05/0, I/088/06/0, and I/008/07/0, 
at PSU by NASA contract NAS5-00136.
H.A.K. was supported by the {\it Swift } project.
P.R.\ thanks INAF-IASF Milano and L.S.\ INAF-IASF Palermo,
for their kind hospitality.
Italian researchers acknowledge the support of Nature (455, 835-836) and thank
the Editors for increasing the international awareness of the current
critical situation of the Italian Research.

\end{document}